\documentclass[aps,prl,twocolumn,a4paper,superscriptaddress,nolongbibliography,10pt]{revtex4-2}

\usepackage{pdfpages}
\usepackage{etoolbox}
\apptocmd{\sloppy}{\hbadness 20000\relax}{}{}

\makeatletter
\AtBeginDocument{\let\LS@rot\@undefined}
\makeatother

\usepackage{epsfig}
\usepackage[makeroom]{cancel} 
\usepackage[normalem]{ulem} 
\usepackage{xcolor} 
\pagecolor{white}
\usepackage{amssymb,graphics,graphicx,amsfonts,amsmath,empheq}
\usepackage[breakable,most,many]{tcolorbox}
\usepackage{varwidth}
\usepackage{capt-of}
\usepackage[export]{adjustbox}
\usepackage{adjustbox}
\usepackage{natbib}
\usepackage{diagbox}
\usepackage{chngcntr}
\usepackage{enumerate}
\usepackage{physics}

\usepackage{hyperref}
\hypersetup{
    colorlinks,
    linkcolor={red!50!black},
    citecolor={blue!50!black},
    urlcolor={blue!80!black}}

\begin{document}

\title{High-Speed Combinatorial Polymerization Through Kinetic-Trap Encoding}

\author{Félix Benoist}
\email{fbenoist@igc.gulbenkian.pt}
\affiliation{Instituto Gulbenkian de Ci\^encia, Oeiras, Portugal}

\author{Pablo Sartori}
\email{psartori@igc.gulbenkian.pt}
\affiliation{Instituto Gulbenkian de Ci\^encia, Oeiras, Portugal}

\begin{abstract}
Like the letters in the alphabet forming words, reusing components of a heterogeneous mixture is an efficient strategy for assembling a large number of target structures. Examples range from synthetic DNA origami to proteins self-assembling into complexes. The standard self-assembly paradigm views target structures as free-energy minima of a mixture. While this is an appealing picture, at high speed structures may be kinetically trapped in local minima, reducing self-assembly accuracy. How then can high speed, high accuracy, and combinatorial usage of components coexist? We propose to reconcile these three concepts not by avoiding kinetic traps, but by exploiting them to encode target structures. This can be achieved by sculpting the kinetic pathways of the mixture, instead of its free-energy landscape. We formalize these ideas in a minimal toy model, for which we analytically estimate the encoding capacity and kinetic characteristics, in agreement with simulations. Our results may be generalized to other soft-matter systems capable of computation, such as liquid mixtures or elastic networks, and pave the way for high-dimensional information processing far from equilibrium.
\end{abstract}
\maketitle

{\bf Introduction.} The combinatorial usage of different components is a prevalent biological strategy to encode information. For example, in the cytoplasm proteins accurately self-assemble into complexes that share proteins with one another~\cite{Gavin06, Kuhner09}. This notion has also permeated nanotechnology \cite{Dunn15,Meng16}, where the same set of DNA tiles can be reused to reliably self-assemble multiple structures~\cite{Evans24}. Besides reusability and high accuracy, a fundamental property of biological self-assembly is high speed, which allows cellular adaptation to quickly changing conditions. This motivates the fundamental question of how self-assembly with reusable components can occur fast and accurately.

The standard approach to combinatorial self-assembly encodes target structures as minima of the mixture's free-energy landscape~\cite{Murugan_PNAS15, Bisker18, Sartori20}. While never explicitly mentioned, this approach is subject to a speed-accuracy trade-off~\cite{Bennett79, Sartori13}: self-assembly of targets is accurate when the mixture can relax to target minima in near-equilibrium conditions. Far from equilibrium, as required for high speed, free-energy encoding results in undesired structures trapping the kinetics. To reconcile self-assembly speed, accuracy, and reusability we propose an alternative encoding approach: tuning the kinetics of the pathways leading to target structures. In this approach, kinetic traps, normally understood as deleterious~\cite{Gartner22, Jia20, Sanz07, Osat24}, can be exploited to encode information that is accessible far from equilibrium and at high speed~\cite{Dill97, Murugan_Nat15, Lefebvre23, Whitelam12}. While tuning the kinetics of different binding partners is a well-established mechanism for discrimination in copolymerization processes~\cite{Sartori13,Tsai06}, its role in self-assembly remains under-studied.

\begin{figure}
	\centerline{\includegraphics[]{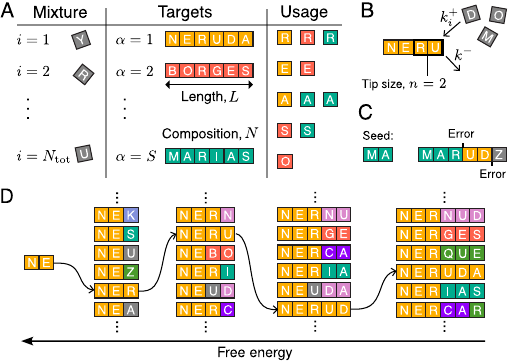}}
	\caption{\label{fig:sketch}{\it Schematics of kinetic encoding setup.} {\bf A.} A heterogeneous mixture of $N_{\rm tot}$ monomer species is designed to self-assemble $S$ different target strings of size $L$ and containing $N$ different components, which results in combinatorial usage of components. {\bf B.} A polymer grows by adding/removing monomers at its tip with rates $k_i^{+}$ and $k^-$ that depend on the composition of the $n$ components of the polymer tip. {\bf C.} When multiple targets are encoded, different nucleation seeds should retrieve different targets. Retrieval is hampered by errors due to the reusability of components (first error) or thermal fluctuations (second error). {\bf D.} In kinetic encoding there is no free energy difference between different strings of the same size. Instead, the pathway to the target is differentiated by kinetics.}
\end{figure}

In this article, we model the dynamics of a self-assembling heteropolymer in contact with a reservoir of multiple different component species. We show how a large number of target strings can be encoded kinetically, such that accurate self-assembly of any of them will occur at high speed. Furthermore, we analytically calculate and numerically confirm the scaling of the maximum number of structures that a mixture can kinetically encode, the characteristic lifetime of targets, and the dependence of our results on the heterogeneity of targets and their usage of components.
\newline

{\bf Model setup.} We consider a mixture of $N_{\rm tot}$ different monomer species, see Fig.~\ref{fig:sketch}A, labeled $i=1, \ldots, N_{\rm tot}$, that are kept at fixed chemical potentials, $\mu_i=\mu$, and temperature, $T$ (hereafter $k_{\rm B}T=1$, with $k_{\rm B}$ Boltzmann's constant). We are interested in conditions under which the mixture can self-assemble any of $S$ different target strings, labeled $\alpha=1,\ldots,S$, that are defined through composition vectors ${\bf c}^{\alpha}=\{i_1^{\alpha}, i_2^{\alpha}, \ldots,i_L^{\alpha}\}$, with $L$ the length (equal for all targets). Each target contains $N\le N_{\rm tot}$ different monomer species, and is thus characterized by its usage of the mixture, $u=N/N_{\rm tot}\le1$, and its compositional heterogeneity, $h=N/L\le1$. The reusability of components, within and across strings, allows for a combinatorial expansion of the mixture.

We study the dynamics of a heteropolymer that grows by adding and removing monomers of the mixture at its distal end, see Fig.~\ref{fig:sketch}B-D. The polymer in question is characterized by its composition vector $\{i_1, i_{2}, \ldots,i_{\ell}\}$, with $\ell$ the time-varying length. The addition and removal of monomers depends on the composition of the polymer tip, ${\bf t}_n=\{i_{\ell+1-n},\ldots,i_\ell\}$, with $n\ge1$ the length of the tip, \textit{i.e.} the range of interactions. We denote the addition rate of a monomer from species $i$ as $k_i^+({\bf t}_{n})$, and the removal rate of monomer $i_\ell$ as $k^-({\bf t}_{n+1})$. Note that the case $n=1$ corresponds to near-neighbor coupling among monomers, $n=2$ to next near-neighbor, etc. Within this setup, our goal is to propose a choice of rates that allows polymerization of target strings reliably and fast.

To ensure fast retrieval of targets, we encode their compositions in the binding kinetics, rather than in the energetics, of the mixture components. Therefore, considering the binding of component $i$ to a tip ${\bf t}_n$ and its subsequent unbinding (from a tip ${\bf t}'_{n+1}$ corresponding to the previous tip ${\bf t}_n$ to which has been added $i$), the detailed balance condition on the rates reduces to
\begin{align}\label{eq:mu}
	k_i^+({\bf t}_{n})/ k^-({\bf t}'_{n+1})=\exp(\mu)\quad,
\end{align}
which encodes no information about the targets. The targets are instead kinetically encoded through the choice of forward rates
\begin{align}\label{eq:d}
	k_i^{+}({\bf t}_{n}) = \exp({r_i\delta})\quad,
\end{align}
where $r_i$ is the number of monomers in the tip ${\bf t}_{n}$ that are correctly placed relative to monomer $i$ at location $\ell + 1$ in any target string ${\bf c}^\alpha$, see SI for explicit formula, and $\delta$ is a kinetic discrimination parameter. Note that the rates are defined up to an irrelevant time unit. As an illustration, for the simple case $n=1$ this rule implies: $r_i=1$, if the monomer $i$ to be added is a neighbor of the tip monomer $i_\ell$ in any of the target strings; and $r_i=0$ otherwise. Alternatively, in the example of Fig.~\ref{fig:sketch}B with $n=2$, we have $k_{\rm D}^+(\{{\rm R,U}\})=\exp(2\delta)$, due to the target ${\bf c}^1=\{{\rm N,E,R,U,D,A}\}$, but $k_{\rm Z}^+(\{{\rm R,U}\})=1$. In the following, we study the conditions under which this minimal model allows accurate and fast retrieval of targets.
\newline

\begin{figure}
	\centerline{\includegraphics[]{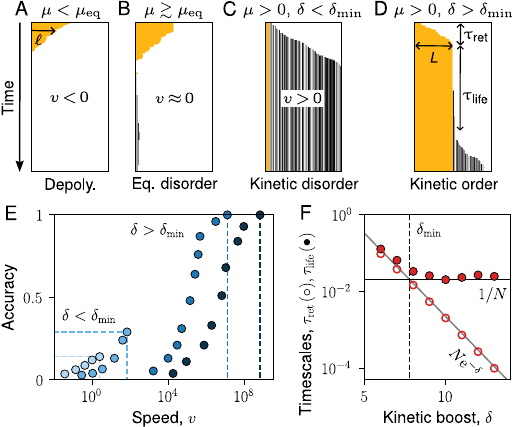}}
	\caption{\label{fig:single}{\it Fast and accurate retrieval of a single target string.} {\bf A-D.} Kymographs of polymer dynamics obtained from stochastic simulations (white codes for empty sites, yellow for components matching the target string, and shades of gray assembly errors), see SI for simulation details. In this and all panels the mixture encodes a single target ($S=1$) that is heterogeneous and uses all components ($L=N_{\rm tot}=N=50$). The polymer tip size is $n=1$ (near-neighbor coupling). Depending on the chemical potential ($\mu$) and discrimination barrier ($\delta$) we identify four different kinetic regimes. {\bf E.} As $\mu$ increases, both the accuracy of target retrieval and the speed increase. The maximal accuracy and maximal speed (dashed lines) for $\mu\to\infty$ both increase with $\delta$. In particular, the accuracy approaches one for $\delta>\delta_{\rm min}$ [Eq.~\eqref{eq:d_min}]. Here, $n=2$ and shades of blue label $\delta=\{4,6,12,14\}$ by increasing darkness. {\bf F.} The timescale of target retrieval ($\tau_{\rm ret}$) and the lifetime of a target ($\tau_{\rm life}$) separate for $\delta>\delta_{\rm min}$, in quantitative agreement with Eq.~\eqref{eq:tau}. Here again $n=1$.}
\end{figure}
 
{\bf Retrieving a target string as a kinetic trap.} As a starting point, we consider that the mixture encodes a single string ($S=1$) that is fully heterogeneous ($h=1$) and uses all components ($u=1$, such that $L=N_{\rm tot}=N$). In this case, errors are not due to combinatorial usage of components (first error in Fig.~\ref{fig:sketch}C), but instead emerge from thermal fluctuations (second error in Fig.~\ref{fig:sketch}C). At equilibrium, the chemical potential of the mixture balances the entropic tendency to grow, and so $\mu_{\rm eq}=-\ln N<0$~\cite{Bennett79, Sartori13, Andrieux08}. Equation~\eqref{eq:mu} implies that no information is encoded in the binding energies, and so the equilibrium state of the polymer is fully disordered: for $\mu\gtrsim\mu_{\rm eq}$ an initially ordered seed will disassemble in favor of a disordered polymer (Fig.~\ref{fig:single}B). This equilibrium state defines a boundary between a depolymerization regime, where the growth speed $v$ (defined as the net rate of monomer addition) is negative, {\it i.e.} $v<0$ for $\mu<\mu_{\rm eq}$ (Fig.~\ref{fig:single}A); and different growth regimes, for which $\mu>\mu_{\rm eq}$ implies $v>0$ (Fig.~\ref{fig:single}C and D).

Equation~\eqref{eq:d} establishes that target strings are encoded in the kinetics, instead of the energetics. Therefore, the accuracy of retrieval should be maximal when the dynamics are strongly irreversible~\cite{Sartori13}. Since in accurate and irreversible dynamics there is only one possible assembly pathway, the bound on the driving is raised to $\mu>0$. Furthermore, suppressing errors due to the presence of $N-1$ confounding monomers at each growth step requires that the kinetic discrimination barrier, $\delta$, be sufficiently large.

To bound $\delta$, we estimate the error rate $p_\text{err}$ as the ratio of the sum of addition rates for all potential erroneous additions to the addition rate of the correct monomer, \textit{i.e.} $p_\text{err}\approx (N-1)/\exp(n\delta)$. In the highly irreversible regime, the probability to retrieve the seeded target can be estimated as $(1-p_\text{err})^{N}$. For $N\gg1$, a significant retrieval probability thus requires $Np_\text{err}\ll1$. This leads to a lower bound on the kinetic discrimination parameter:
\begin{equation}\label{eq:d_min}
	\delta_{\rm min} = \frac2n\ln N \quad.
\end{equation}
We distinguish two fast-growth regimes: kinetic disorder for $\delta<\delta_{\rm min}$, in which thermal fluctuations result in frequent addition errors, \textit{i.e.} $p_\text{err}\approx1$ (Fig.~\ref{fig:single}C); and kinetic order for $\delta>\delta_{\rm min}$, in which a target string is accurately retrieved, \textit{i.e.} $p_\text{err}\approx0$, until a fluctuation destabilizes it (Fig.~\ref{fig:single}D).

Figure~\ref{fig:single}E shows that increasing the driving $\mu$ results in an increase of the growth speed, up to saturation at $v\approx \exp(n\delta)$, as well as an increase in retrieval accuracy (defined as the fraction of string length assembled until the first error). Still, high accuracy is only possible for large discrimination barriers, in agreement with Eq.~\eqref{eq:d_min}.

Kinetic encoding implies that targets are not thermodynamically stable. We can however estimate their kinetic stability. The time it takes to retrieve a target string, $\tau_{\rm ret}$, is obtained by dividing the length of the string, $N$, by its growth speed, $v$, and so $\tau_{\rm ret}\approx N\exp(-n\delta)$. In contrast, the lifetime of the string, $\tau_{\rm life}$, is given by the time it takes to add a few incorrect monomers, and so $\tau_{\rm life}\approx1/N$. Therefore, the lifetime of a string relative to its retrieval time reads
\begin{align}\label{eq:tau}
	\tau_{\rm life}/\tau_{\rm ret}\approx\exp(n\delta)/N^2 \quad,
\end{align}
and so larger discrimination barriers and longer tip sizes result in more stable strings, see Fig.~\ref{fig:single}F. To conclude, we have shown that the kinetic encoding approach in Eq.~\eqref{eq:d} allows for fast and accurate retrieval of a single target string for strong discrimination far from equilibrium.
\newline

\begin{figure}
	\centerline{\includegraphics[]{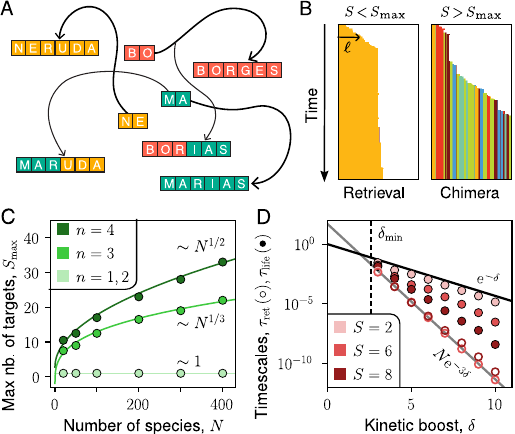}}
	\caption{\label{fig:scaling}{\it Combinatorial encoding of kinetic pathways.} {\bf A.} Reusing components across targets strings can result in chimeric assemblies, as kinetic pathways will interfere (a kinetic analogy of cross-talk of multiple minima). {\bf B.} To avoid chimeric strings, the number of target strings $S$ must be smaller than $S_{\rm max}$ [Eq.~\eqref{eq:Smax1}]. Here $S_\text{max}\approx3$, and the two values of $S$ are 2 and 6. Throughout this figure we used $N=L=N_\text{tot}=50$ and $\mu=3$. In this panel, $\delta=5$ and $n=2$. {\bf C.} The capacity limit increases with the target size $N$ depending on the monomer connectivity $n$ according to Eq.~\eqref{eq:Smax1}. {\bf D.} Target stability deteriorates with increasing number of target strings $S$. Here, $n=3$, such that $S_\text{max}=9$.}
\end{figure}

{\bf Combinatorial encoding far from equilibrium.} We now turn to the case in which the mixture encodes multiple targets ($S>1$) by combinatorially reusing components across targets~\cite{Murugan_PNAS15, Sartori20}. In this scenario, errors arise when one component has as neighbors two different components in two different targets, making these two targets indistinguishable to the tip of a growing polymer. For example, in Fig.~\ref{fig:sketch}C for the state $\{{\rm M,A,R}\}$ of the polymer and $n=1$ we have that $k_{\rm I}^+=k_{\rm U}^+=\exp(\delta)$, due to $\{{\rm R}\}$ appearing both in ${\bf c}^1=\{{\rm N,E,R,U,D,A}\}$ and ${\bf c}^S=\{{\rm M,A,R,I,A,S}\}$, which can result in the error shown. Such types of errors hence cannot be suppressed by increasing $\delta$. Conceptually, if the mixture encodes many kinetic pathways to different targets, such pathways may cross, making it likely to retrieve a chimera (formed by fragments of different target strings) rather than the seeded target string (Fig.~\ref{fig:scaling}A). In Fig.~\ref{fig:scaling}B, we show two examples of successful and failed retrieval for a mixture that kinetically encodes multiple target strings. For the same large positive values of $\mu$ and $\delta$, if the number of stored targets is below a certain maximal value, $S<S_{\rm max}$, retrieval is successful; instead if $S>S_{\rm max}$, the initial seed nucleated fragments from many other different target strings, yielding a chimeric polymer~\cite{Murugan_PNAS15, Sartori20}.

What determines the maximum number of target strings, $S_{\rm max}$, that can be accurately assembled from a mixture via kinetic encoding? To answer this question, we define the promiscuity of components, $\pi$, as the number of specific interactions of a typical component at each near-neighboring location. For instance, for the targets shown in Fig.~\ref{fig:sketch}A, monomer $\{\text R\}$ interacts with $\pi=3$ different monomer species at each near-neighbor location. A large promiscuity turns components indistinguishable, irrespective of $\delta$, hampering the reliability of assembly. The error rate, $p_{\rm err}$, thus denotes the probability that given a tip of size $n$ there is ambiguity regarding which component can be added. For $n=1$, we can estimate $p_\text{err}\approx(\pi-1)/\pi$, whereas for $n\ge2$, the error rate scales as $p_\text{err}\sim(\pi-1)^n/N^{n-1}$ (see SI for derivation). We can then proceed to satisfy the condition $Np_\text{err}\ll1$ as in the derivation of Eq.~\eqref{eq:d_min}, focusing on the case where all components are used once in every target ($h=u=1$), for which $\pi\approx S$. We obtain $S_{\rm max}=1$ for $n=1$, because the error rate, $p_\text{err}\approx(S-1)/S$, prevents retrieval for as little as two targets; whereas for $n\ge2$,
\begin{equation}\label{eq:Smax1}
	S_{\rm max}\sim N^{1-2/n} \quad.
\end{equation}
The predicted size scaling goes from $\order{1}$ for $n=2$, and thus no combinatorial usage is possible, to $\order{N}$ for the fully-connected case $n=N$, akin to neural network capacity~\cite{Hopfield82}. Therefore, increasing the tip size improves discrimination, which allows to encode more targets.

Figure~\ref{fig:scaling}C shows the results of numerical simulations relating the capacity, $S_{\rm max}$, to the number of component species, $N$, for different tip sizes, $n$. As one can see, for $n=1,2$ no combinatorial usage of components is possible, whereas for $n=3,4$ the numerical results are in good agreement with the predictions of Eq.~\eqref{eq:Smax1}. Figure~\ref{fig:scaling}D shows how the time of retrieval, $\tau_\text{ret}$, and lifetime, $\tau_{\rm life}$ depend on the kinetic discrimination parameter $\delta$ for different numbers of target strings, $S$. While $\tau_\text{ret}$ follows the behavior derived earlier, $\tau_\text{life}$ is now bounded by $\exp(-\delta)$, because errors are dominated by component reusability. As $S$ increases, $\tau_\text{life}$ decreases, making structures more unstable as $S$ approaches the capacity limit, $S_{\rm max}$. Overall, we conclude that kinetic encoding of a combinatorially large number of components is possible, with a capacity and stability that agree with our analytical estimates.
\newline 

\begin{figure}
\centerline{\includegraphics[]{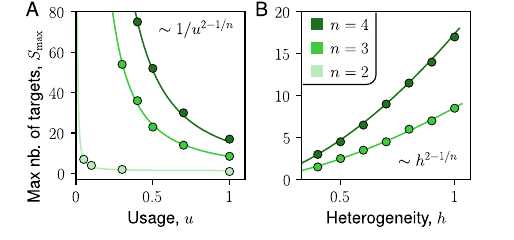}}
\caption{\label{fig:hetspar}{\it Effects of component usage and heterogeneity in combinatorial encoding.} {\bf A.} The maximum capacity, $S_{\rm max}$, decreases with increasing usage, $u$, at fixed length $L$ as predicted by Eq.~\eqref{eq:Smax2}. {\bf B.} $S_{\rm max}$ increases with increasing heterogeneity, $h$, also in quantitative agreement with Eq.~\eqref{eq:Smax2}. In both panels, $L=25$ for $n=2$, and $L=100$ for $n=3,4$.}
\end{figure}

{\bf The roles of heterogeneity and usage.} We now study the effect of target heterogeneity, $h$, and usage of components, $u$, on the capacity of the system, $S_{\rm max}$. In this general case, the promiscuity of components is given by $\pi\approx Su/h$ for large heterogeneous targets [SI]. 
Following an argument analogous to the previous section, {\it i.e.} $Lp_\text{err}\ll1$ with $p_\text{err}\sim(\pi-1)^n/N_{\rm tot}^{n-1}$, the scaling in Eq.~\eqref{eq:Smax1} generalizes to 
\begin{align}\label{eq:Smax2}
	S_{\rm max}\sim(h/u)^{2-1/n}L^{1-2/n} \quad,
\end{align}
for $n\ge2$. This expression highlights the role of the target string length $L$ as a key extensive quantity regulating the scaling. Eq.~\eqref{eq:Smax2} also shows that increasing heterogeneity and reducing usage both result in an increase of the capacity. The intuition behind this is simple. Increasing heterogeneity will reduce the reusability of components within each target. Likewise, reducing the usage of components made by each target string reduces the reusability across targets. Both aspects reduce the promiscuity of components, thus facilitating that polymerization pathways of different targets stay separate from each other. Figure~\ref{fig:hetspar} shows that our numerical simulations recover the capacity scaling in Eq.~\eqref{eq:Smax2}. In particular the capacity diverges as the usage becomes low ($u\to0$), and reaches a maximum for fully heterogeneous structures ($h=1$). Therefore, high heterogeneity and low usage of available components results in increased capacity for kinetic encoding, as dictated by Eq.~\eqref{eq:Smax2}.
\newline

{\bf Discussion.} Classical self-assembly relies on the stability of target structures, which results in a trade-off between self-assembly speed and accuracy. In contrast, combining ideas of information thermodynamics~\cite{Bennett79, Sartori13, Bennett82} and neuroscience~\cite{Hopfield82, Hertz91}, we have shown how self-assembly of heterogeneous target structures can be performed kinetically. In this approach, higher speed implies higher accuracy, breaking the aforementioned trade-off~\cite{Pigolotti16,Banerjee17}. Since assemblies in this scenario do not correspond to deep energy minima, but to long-lived kinetic traps, they are only stable for a finite amount of time.

The concept of kinetic encoding of target structures is motivated by the self-assembly of large heteromeric protein complexes, such as the ribosome~\cite{Klinge19}. It is well established that specific binding events for these systems are catalyzed by enzymes lowering the kinetic barrier for binding~\cite{Davis17,Dorner23}. The importance of such enzymes, often referred to as assembly factors, cannot be understated, as their deletion results in the slow-down of the self-assembly speed and the accumulation of incomplete structures~\cite{Seffouh24,Davis17}. In fact, there are often as many assembly factors as components in a given assembly~\cite{Dorner23,Wahl09,Vercellino22}. In this sense, our choice of binding rates can be understood as coarse-graining enzymatic reactions due to assembly factors. More broadly, both kinetic encoding (as studied here) and energetic encoding (see Refs.~\cite{Bisker18,Murugan_PNAS15,Sartori20}) are expected to play a role. How best to combine them for maximal efficiency remains an open question, see SI for additional discussion details. 

While we have illustrated the concept of kinetic-trap encoding in a toy model of heteromeric polymerization, this idea may be adaptable to systems in other branches of soft-matter physics, which at present all use energetic interactions to encode target functions. Examples include self-assembly of structures with more complex geometries~\cite{Lenz_Nat17,Tyukodi22}, programmable liquid phases~\cite{Teixeira24,Zwicker22}, colloidal self-assembly~\cite{McMullen22, Romano20}, guided self-folding~\cite{Dunn15,Pinto24}, or elastic network models of proteins~\cite{Yan17, Rocks17}. Since the biophysical systems that these models aim to describe often operate far from equilibrium, we expect the generic features of kinetic encoding here presented will play a central role in explaining how biological matter is capable of complex high-dimensional information processing.
\newline

\acknowledgments{This work was partly supported by a laCaixa Foundation grant (LCF/BQ/PI21/11830032) and core funding from the Gulbenkian Foundation.}


%

\pagebreak

$ $

\newpage
\widetext
\begin{center}
\textbf{\Large Supporting information for ``High-Speed Combinatorial Polymerization \\\vspace{1.5mm} Through Kinetic-Trap Encoding''}
\end{center}
\setcounter{section}{0}
\setcounter{equation}{0}
\setcounter{figure}{0}
\setcounter{table}{0}
\setcounter{page}{1}
\makeatletter
\renewcommand{\theequation}{S\arabic{equation}}
\renewcommand{\thefigure}{S\arabic{figure}}
\vspace{1cm}

{\bf Simulation details.} To model the growth process, we use the Gillespie algorithm. Starting from a nucleation seed, each iteration consists of three tasks. (i) Based on the tip $\mathbf t_n$, we consider the rates of adding a monomer from any species, $k_i^+(\mathbf t_n),\ i=1,\dots,N_{\rm tot}$, and the rate of removing the last monomer bound, $k_-(\mathbf t_n)$, and we compute their sum which we name $k_{\rm out}$, based on Eqs.~(\ref{eq:mu}-\ref{eq:d}). The calculation of each rate involves comparison with targets components from each of the $S$ targets over the tip with $n$ neighboring locations. (ii) The waiting time is drawn from an exponential distribution with mean $k_{\rm out}^{-1}$. (iii) A move $\mu$, referring to the adding of a monomer or the removing of the last monomer, with rate $k$ is chosen with the probability $k_\mu/k_{\rm out}$. Overall, each step requires a time complexity of $\order{nSN_{\rm tot}^2}$.
\newline

{\bf Encoding details.} Each target structure labeled $\alpha=1,\dots,S$ is encoded through a target composition vector $\mathbf c^\alpha$. Due to reusage of components within and across strings, we label the instances of species $i$ in all targets as $\beta=1,\dots,B_i$ with $B_i\le SL$. Component $i$ is thus associated with a set of encoded neighborhoods $\mathbf n^\beta(i)=\{i_{-n}^\beta,\dots,i_{-1}^\beta\}$, each with size $n$, corresponding to subsets of the target composition vectors $\mathbf c^\alpha$. For components close to the left-most target boundary, the missing neighbors are written with $-1$. The number $r_i$ characterizing the binding and unbinding of monomer $i$ at location $l+1$ of a growing polymer thus equals the number of neighbors in $\mathbf t_n=\{i_{\ell+1-n},\ldots,i_\ell\}$ that are also in any target neighborhood $\mathbf n^\beta(i)$ at the corresponding location. The missing neighbors in $\mathbf t_n$ are written with $-2$. This is such that, \emph{e.g.}, neighbor $i_l$ is checked for correspondence with all encoded neighbors $i_{-1}^1,\dots,i_{-1}^{B_i}$. Using the Heaviside function $\Theta$, we can mathematically write
\begin{equation}\label{eq:r_ix}
	r_i=\sum_{y=1}^n\Theta\bigg[\sum_{\beta=1}^{B_i}f(i_{l+1-y},i_{-y}^\beta)\bigg], \qq{where} f(x,y)=\delta_{xy}.
\end{equation}

\begin{figure} [!b]
\centerline{\includegraphics[width=.6\linewidth]{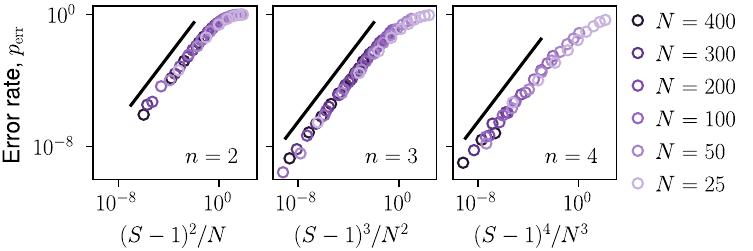}}
\caption{\label{fig:error}{\it The error rate follows the prediction.} In the regime $h=u=1$, where the promiscuity $\pi\approx S$, the prediction for the error rate $p_{\rm err}\sim (S-1)^{n}/N^{n-1}$ agrees well with numerical calculations. In particular, the data at different sizes $N$ all collapse. Here, $p_{\rm err}$ is averaged over the $L$ assembly steps. The black line has slope 1.}
\end{figure}

{\bf \textit{Ansatz} for the error rate.}
In the regime where the promiscuity $\pi$ is large, components become indistinguishable, irrespective of $\delta$, which prevents reliable assembly. The error rate, $p_{\rm err}$, defined as the probability that there is ambiguity regarding which component can be added, is estimated as follows in the general case $u,h\le1$. For any given tip size $n$, an error requires that two components share the same $n$ neighbors either in different targets or in different locations in the same target. Due to component reusage, the error rate for $n=1$ is simply proportional to the number of specific neighbors, except the one continuing the string in formation: $(\pi-1)$. Then, sharing an additional neighbor should be decreased by a factor $(\pi-1)/N_\text{tot}$, due to the large number of species. For large $\delta$, we thus expect that the average error rate is 
\begin{equation}
	p_\text{err}\sim(\pi-1)\left(\frac{\pi-1}{N_\text{tot}}\right)^{n-1}. 
\end{equation}
This estimate is numerically validated in Fig.~\ref{fig:error} for $n\ge2$ and $N_{\rm tot}$ large, where it is below 1. The agreement is particularly good at low error rate, where the estimate yields a large capacity $S_{\rm max}$. For $n=1$, a better estimate is $p_\text{err}\approx(\pi-1)/\pi$ simply counting the fraction of specific neighbors that leads to a retrieval error. 

\newpage

\begin{figure} [!h]
\centerline{\includegraphics[width=.5\linewidth]{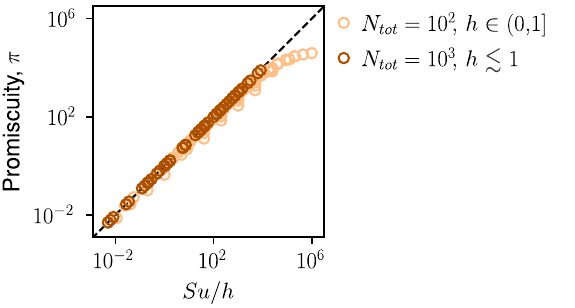}}
\caption{\label{fig:prom}{\it The promiscuity follows the prediction.} We check the prediction for the promiscuity $\pi\approx Su/h$ with numerical calculations. In the full range $h\in(0,1]$ and $u\in(0,1]$ and for modest sizes $N_{\rm tot}=10^2$ (orange circles), the agreement is good. And for $h$ close to 1 and larger sizes $N_{\rm tot}=10^3$ (brown circles), the agreement is great. The promiscuity is averaged over the $N_\text{tot}$ components.}
\end{figure}

{\bf \textit{Ansatz} for the promiscuity.} In practice, we calculate the promiscuity $\pi_i$ of species $i$ by counting the number of different encoded right neighbors across all instances of species $i$ in all targets labeled $\beta=1,\dots,B_i$. Given the encoded neighborhoods $\mathbf n^\beta(i)=\{i_{-n}^\beta,\dots,i_{-1}^\beta\}$, this corresponds mathematically to the size of the set $\{i_{-1}^1,\dots,i_{-1}^{B_i}\}$, inferior to $SL$, having removed all redundancies. The promiscuity $\pi$ as defined in the text corresponds to the average of the $\pi_i$ across all species $i=1,\dots,N_{\rm tot}$. In the general case $u,h\le1$, we estimate $\pi\approx Su/h$ for large heterogeneous targets~\cite{Sartori20}. This prediction agrees well with numerical calculations, especially for large heterogeneous targets (brown circles), see Fig.~\ref{fig:prom}.

{\bf Extended discussion.} This section investigates the practical feasibility of implementing our kinetic encoding within physical systems. In self-assembly processes where the on-rates are governed by monomer diffusion, the prospect of kinetic discrimination may seem implausible. However, numerous examples exist, including certain driven systems, where kinetics play a pivotal role, and the on-rates are not solely constrained by diffusion. Below, we examine several examples from chemistry and biology where on-rates are influenced by the state of a system, thereby contextualizing our approach.
\newline

\begin{figure} [!b]
\centerline{\includegraphics[]{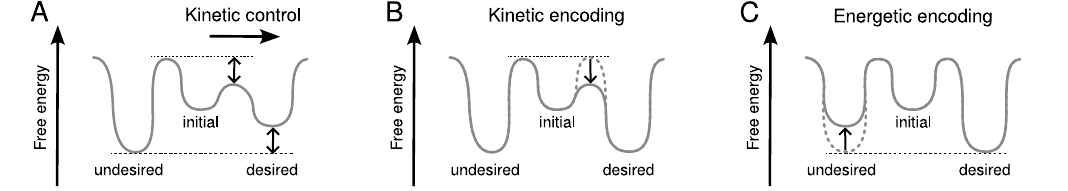}}
\caption{\label{fig:draw} \textit{Examples of different energy landscapes to obtain desired products.} {	\bf A.} Kinetic control in organic chemistry is an example of a situation in which kinetic barriers play a key role in determining the desired product. {\bf B.} Kinetic encoding fully separates differences in energy from difference in kinetic barriers, so the bias in a chemical reaction is purely kinetic. {\bf C.} Energetic encoding removes any difference in the on-rates, so that the difference in the products arises from differences in energy.}
\end{figure}

\textit{From organic chemistry to enzymatic reactions.} On-rates play a critical role in determining the outcomes of chemical reactions. While it is generally assumed that reactions favor products with lower free energy, there are notable cases in organic chemistry where the favored products are determined by the height of the kinetic barrier, see Fig.~\ref{fig:draw}A. This phenomenon, commonly referred to as ``kinetic control''~\cite{ fox2004organic} is typically associated with short reaction times and has significant applications, such as in chemical self-assembly~\cite{Numata18}. Kinetic control thus gives a foundational basis to our setting.

The on-rates of biological processes play a fundamental role in multiple scenarios. One class of such processes is biological polymerization. For example, during the polymerization of amino acid chains by the ribosome, there are intermediate steps wherein the ribosome kinetically discriminates among different tRNA assemblies, associated to conformational changes within the ribosome structure~\cite{Johansson08, gromadski2004kinetic}. Another example is that of DNA replication by polymerases~\cite{Tsai06}, which also perform kinetic discrimination. Besides polymerization processes, kinetics is also fundamental in signaling networks. The simplest example is that of two-component systems in bacteria, where the phosphorylation kinetics between histidine kinases and response regulators are key in determining the response to stimuli~\cite{laub2007specificity}. These examples, beyond the domain of pure chemistry, show how common kinetic discrimination is inside cells.

While certain cellular reactions can occur spontaneously, many reactions are specifically catalyzed by enzymes. As schematically depicted in Fig.~\ref{fig:draw}B, enzymes work by lowering the kinetic barrier, \textit{i.e.} accelerating the on-rate, along the target pathway. The use of enzymes thus constitutes a generic framework for implementing kinetic discrimination~\cite{Alberts15}. There are many examples of enzymatic reactions occurring inside cells, such as those relating to self-assembly of protein complexes, which we discuss further below.
\newline

\textit{Self-assembly of protein complexes.} One of our main motivations is the self-assembly of large heteromeric protein complexes, such as the ribosome. Ribosomes involve about 100 different components~\cite{Klinge19}, which must self-assemble fast and reliably to sustain protein production. The high speed and accuracy of ribosomal assembly are clear indicators that this process is under some form of kinetic encoding, far beyond equilibrium relaxation to a free-energy minimum. \textit{In vivo} ribosomal assembly is aided by numerous assembly factors -- enzymatic proteins that facilitate component binding but are not part of the final ribosome structure. One specific function of assembly factors is that of resolving competition between two components for the same binding site on a growing ribosome complex. By blocking the reactive site of the undesired monomer, they raise its kinetic barrier and promote the binding of the desired monomer~\cite{Davis17}. The high number of assembly factors suggests that they regulate the binding of most components~\cite{Dorner23}. Furthermore, many assembly factors are in fact ATPases, which are strongly driven out of equilibrium. Assembly factors assist the self-assembly of other protein complexes, such as the spliceosome~\cite{Wahl09} or complex I~\cite{Vercellino22}.

Our model for out-of-equilibrium heteromeric polymerization represents a simplified framework for protein-complex assembly. The kinetics of the on-rates can thus be understood as coarse-graining more complex reactions, as has been done for templated polymerization~\cite{Bennett79,Sartori13}. This coarse-graining represents both potential conformational changes of the components and the role of assembly factors. Our toy model especially neglects aspects of energetic encoding, as those were already treated in Ref.~\cite{Sartori20} based on the alternative assumption that self-assembly occurs quasi-statically towards a free-energy minimum, see Fig.~\ref{fig:draw}C. Future research should integrate kinetic and energetic encoding to provide a more realistic model of protein-complex self-assembly.
\newline

Overall, the notion of kinetic discrimination through on-rates is widespread both in the chemical and biological literature, and it is of particular relevance for the self-assembly of, \emph{e.g.}, protein complexes.

\end{document}